\documentclass{article}
\usepackage{arxiv}
\usepackage{times}
\usepackage{cite}
\usepackage{url}
\usepackage[pdftex]{graphicx}
\graphicspath{{Images-DDosDet/}}
\usepackage{subcaption}
\usepackage{rotating}
\usepackage{rotfloat}
\usepackage{xcolor}
\usepackage{amsmath}
\usepackage{amssymb}
\usepackage[linesnumbered,ruled,vlined]{algorithm2e}
\usepackage{pseudocode}
\usepackage[english]{babel}
\usepackage{tabularx}
\usepackage{gensymb}
\usepackage{textcomp}
\usepackage{placeins}
\usepackage{balance}
\usepackage{booktabs}
\usepackage[utf8]{inputenc} 
\usepackage[T1]{fontenc}    
\usepackage{hyperref}       
\usepackage{url}            
\usepackage{booktabs}       
\usepackage{amsfonts}       
\usepackage{nicefrac}       
\usepackage{microtype}      
\usepackage{lipsum}		
\usepackage{graphicx}
\usepackage{natbib}
\usepackage{doi}

\title{DDoSDet: An approach to Detect DDoS attacks using Neural Networks}

\author{ \href{https://orcid.org/
0000-0003-1362-0498}{\includegraphics[scale=0.06]{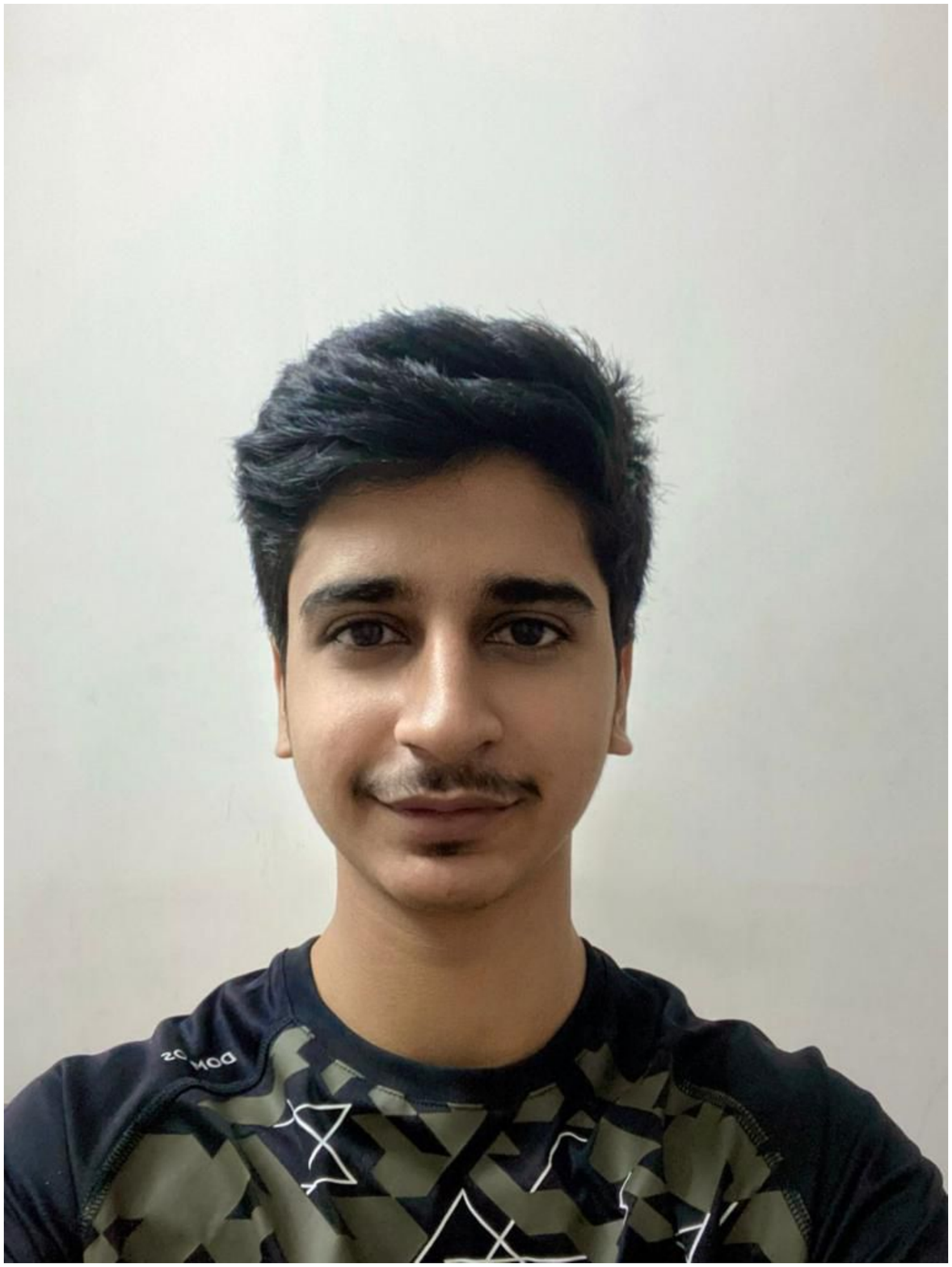}\hspace{1mm}Aman Rangapur}
	\\ School of Computer Science and Engineering\\
	VIT-AP university\\
	Andhra Pradesh, India \\
	\texttt{amanrangapur@gmail.com} \\
	\And
\href{}{\includegraphics[scale=0.06]{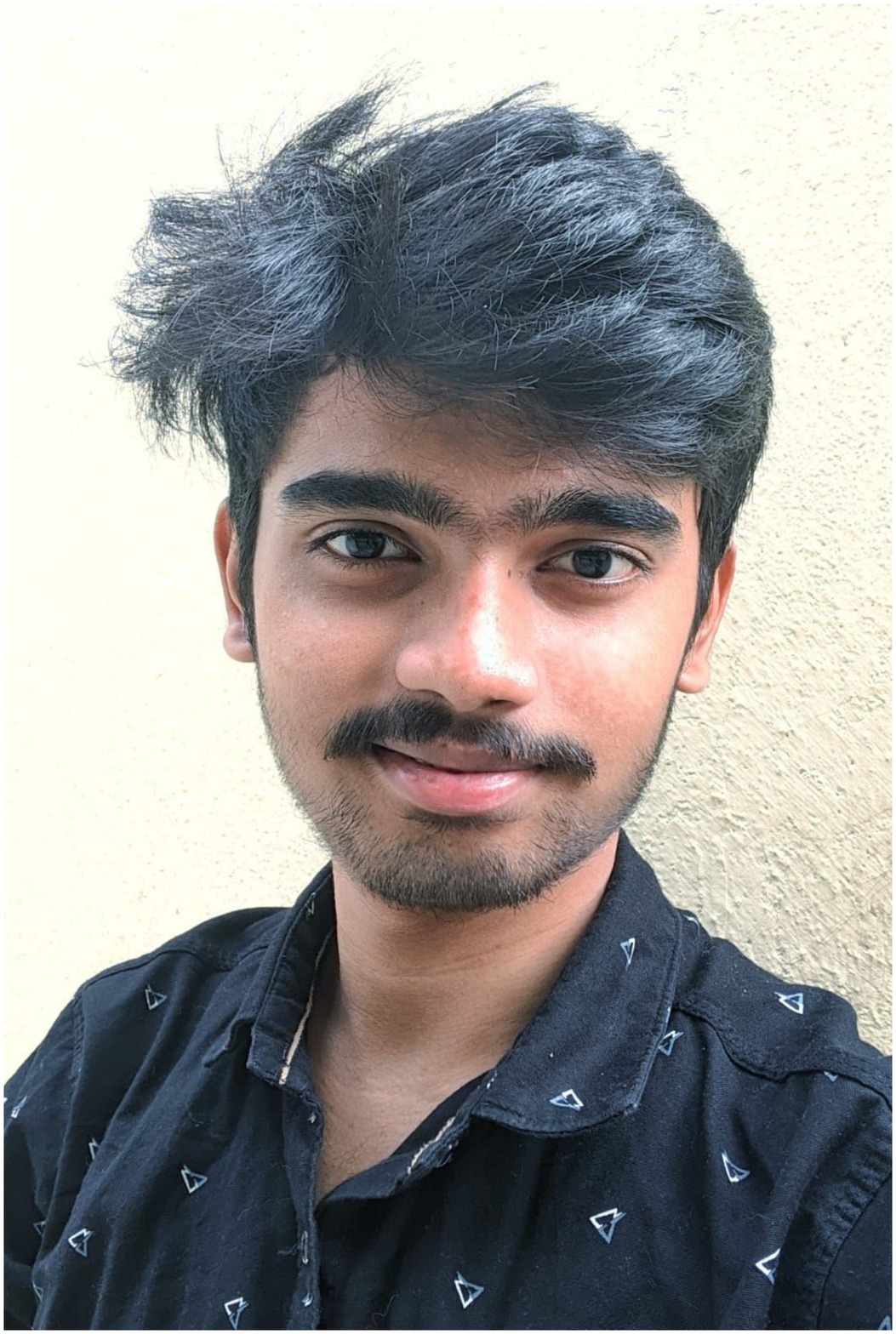}\hspace{1mm}Tarun Kanakam}
        \\
	School of Computer Science and Engineering\\
	VIT-AP university\\
	Andhra Pradesh, India \\
	\texttt{kanakamtarunkumartk@gmail.com} \\
\And
	\href{}{\includegraphics[scale=0.06]{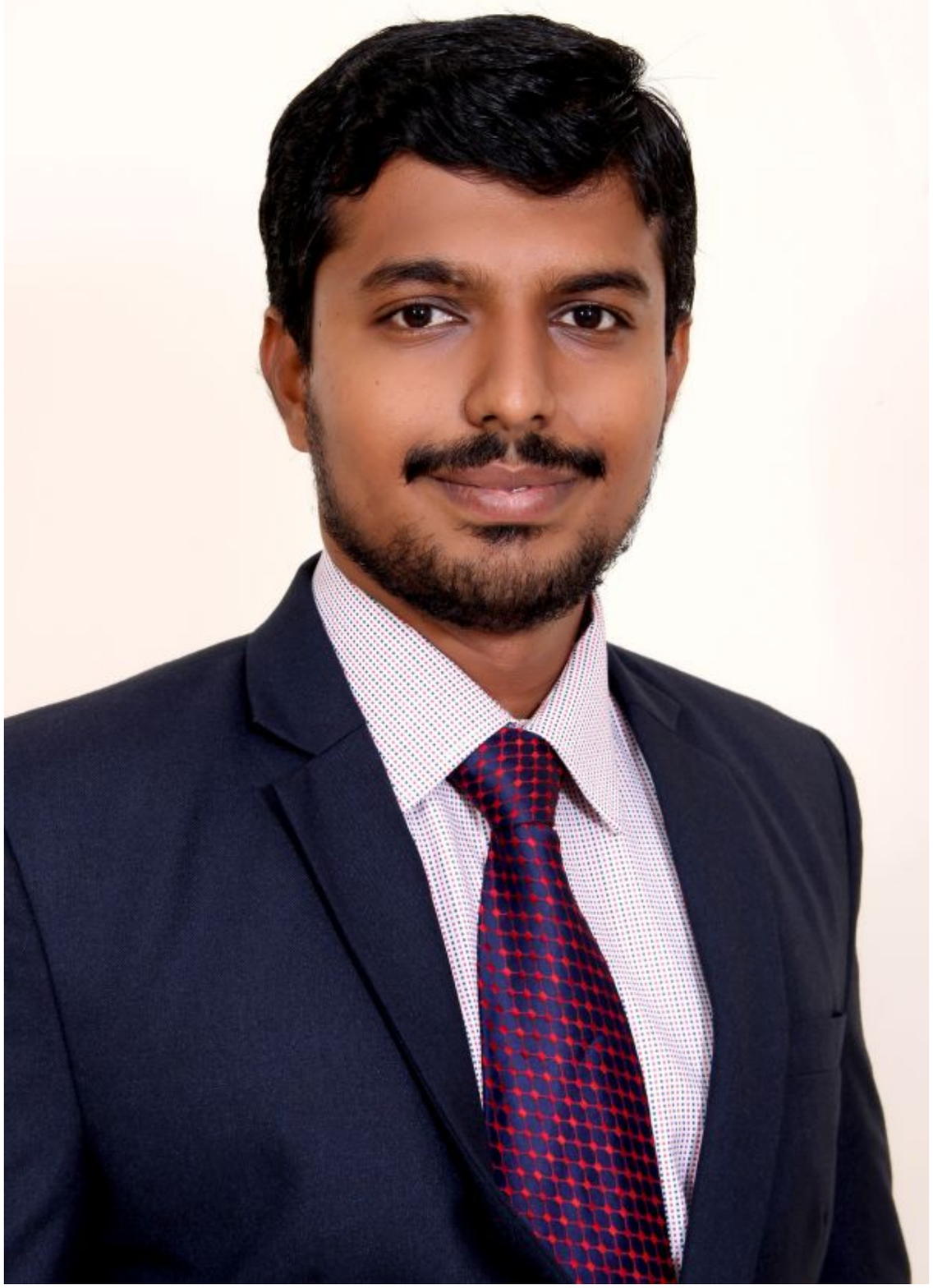}\hspace{1mm}Ajith Jubilson}
	\\
	School of Computer Science and Engineering\\
	VIT-AP university\\
	Andhra Pradesh, India \\
	\texttt{ajith.jubilson@vitap.ac.in} \\
}

\renewcommand{\shorttitle}{\textit{DDoSDet: An approach to Detect DDoS attacks using Neural Networks}}

\hypersetup{
pdftitle={DDoSDet: An approach to Detect DDoS attacks using Neural Networks},
pdfsubject={Artificial Intelligence and Machine Learning},
pdfauthor={Aman Rangapur, Tarun Kanakam and Ajith Jubilson},
pdfkeywords={Distributed Denial of Services (DDoS), Denial of Service (DoS), Intrusion Detection System(IDS), Neural networks.}
}

\begin{document}
\maketitle

\begin{abstract}

Cyber-attacks have been one of the deadliest attacks in today's world. One of them is DDoS (Distributed Denial of Services). It is a cyber-attack in which the attacker attacks and makes a network or a machine unavailable to its intended users temporarily or indefinitely, interrupting services of the host that are connected to a network. To define it in simple terms, It's an attack accomplished by flooding the target machine with unnecessary requests in an attempt to overload and make the systems crash and make the users unable to use that network or a machine. In this research paper, we present the detection of DDoS attacks using neural networks, that would flag malicious and legitimate data flow, preventing network performance degradation. We compared and assessed our suggested system against current models in the field. We are glad to note that our work was 99.7\% accurate.

\end{abstract}

\section{Introduction}
\label{Sec:Introduction}

This modern world is suffering from issues regarding cybersecurity and privacy. It is truly difficult and economically unfeasible to create and maintain such systems, as well as to assure that both the network and the accompanying systems are not vulnerable to threats and assaults. Over the last several decades, there has been a surge in the number of illegal acts in networks, in addition to an increase in devious and malicious content\cite{article1}. When an individual or an organization intentionally and maliciously attempts to enter the information system of another individual or organization, this is referred to as a cyberattack. While most assaults have an economic aim, various recent operations have included data destruction as a goal. Cybersecurity is the need of today's time. Cybersecurity can be defined as the protection of systems, networks, and data within cyberspace\cite{whatiscybersecurity?}.

\begin{figure}
\centering
\includegraphics[width=0.5\textwidth]{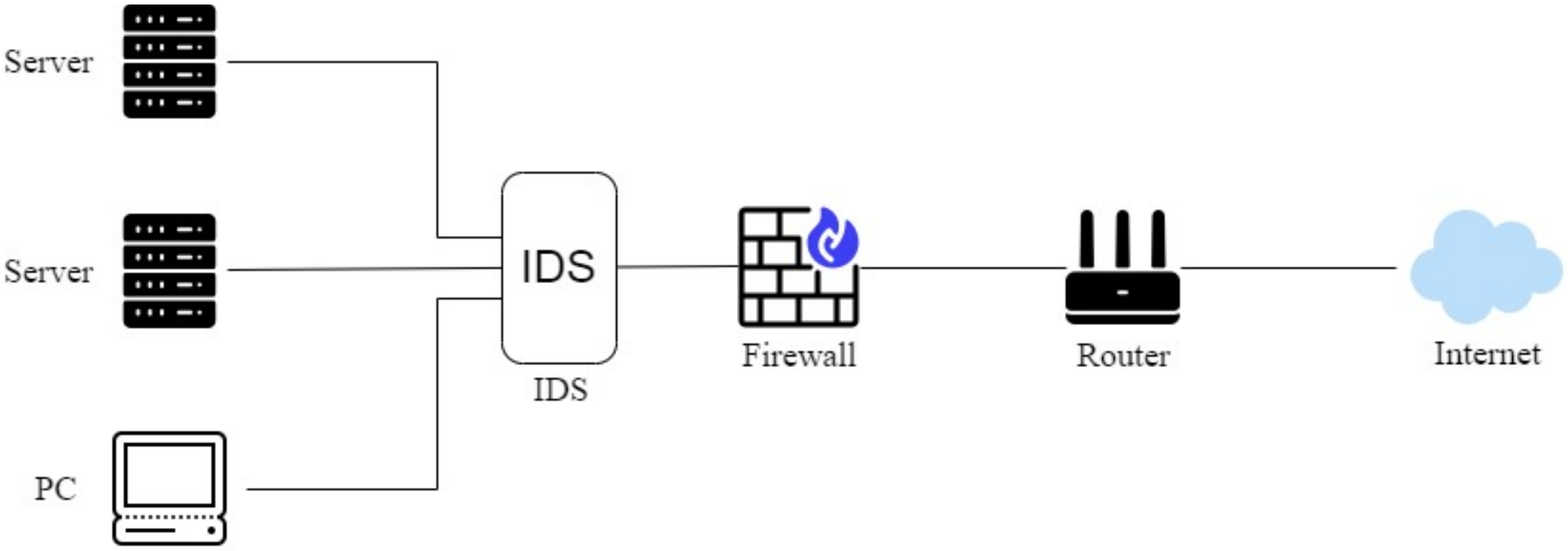}
\caption{Architecture of Intrusion Detection System (IDS)}
\label{FIG:IDS_converted}
\end{figure}

DoS (Denial of Services) attacks, overload resources and bandwidth by flooding systems, servers, and/or networks with traffic\cite{ddos1}. IDS is a powerful tool that is used by network experts to secure data from network attacks that come from different sources\cite{article2}. Any malicious venture or violation is normally reported either to an administrator or collected centrally using a security information and event management (SIEM) system. IDS emerged as a basic and important tool that looks into network security. Fig. \ref{FIG:IDS_converted} explains the architecture of IDS.

The data packets received from the internet are passed to the processing unit, where the data format is altered to make it compatible with the associated IDS, and the data packets are finally classified as an attack or normal\cite{ids1}. Normal data packets are permitted to pass through, while attack data packets are detected as attack types and are stored in the attack database, triggering an alarm and invoking the defensive method\cite{inproceedings1}. A large amount of research is conducted in improving the IDS using Neural Networks. Artificial neural networks have shown to be more advantageous since they need precise, and extensive training, validation, and top-level testing before being deployed to networks to identify fraudulent data and network attacks.

Thus, the main goal of our research is to explore the DoS flooding attack and make a feasible attempt to detect DDoS attacks. This paper presents an overview of the DDoS problem, existing approaches to detect DDoS attacks, and our noble proposal to detect these attacks. Hence, this study suggests a model which is effective and efficient towards detecting DDoS attacks to ensure the safety of the networks.

\section{Background Study}
\label{Sec:Background_Study}
A Distributed Denial of Service (DDoS) attack is defined as an occurrence in which a legitimate person or business is denied access to services that they would ordinarily anticipate, such as web, email, or network connectivity. DDoS is essentially a problem of resource overuse. A DoS or DDoS attack is similar to a large crowd of people crowding at a shop's entry door, making it difficult for genuine customers to enter and disrupting the business\cite{article8}. Bandwidth, memory, CPU cycles, file descriptors, and buffers are all examples of resources. The malicious resource is bombarded by attackers using a flood of packets or a single logic packet that can initiate a cascade of activities to drain the restricted resource\cite{article9}. In a typical DDoS assault, the assailant takes advantage of a weakness in one computer system and turns it into the DDoS master. The assault master system finds additional susceptible systems and takes control of them by infecting them with malware or circumventing authentication controls by guessing the default password of a commonly used system or device.

From 2000 to 2004 a high frequency of DDoS attacks was faced by every organization. The DDoS attack was first reported by Computer Incident Advisory Capability (CIAC) somewhere around the summer of 1999 \cite{canada11}. For example, the greatest DDoS assault on a coin exchange was reported on April 4, 2013, at Mt. Gox, a Tokyo-based corporation, where the price of virtual currency was manipulated and produced a fluctuation with unstable pricing. Additionally, the adversary displayed the traders the error pages\cite{tech13} \cite{kitten14}. Meanwhile, a new attack on Spamhaus has wreaked havoc on the UK and Swiss-based charity organizations. This is one of the world's largest DDoS cyber-attack, with 300 Gigabits per second of data under threat\cite{article15}. Furthermore, At the end of August 2021, Microsoft announced its Azure cloud service mitigated 2.4 terabits per second distributed denial of service assault, the company's greatest DDoS attack to date and the second-largest DDoS attack ever recorded\cite{the16}. The record-breaking DDoS assault occurred in three short waves over ten minutes, with the first at 2.4 Tbps, the second at 0.55 Tbps, and the third at 1.7 Tbps, according to the Microsoft executive. According to Qrator Labs, the operators of the Meris botnet launched a DDoS attack of 21.8 million requests per second in early September.

\begin{figure}
\centering
\includegraphics[width=0.5\textwidth]{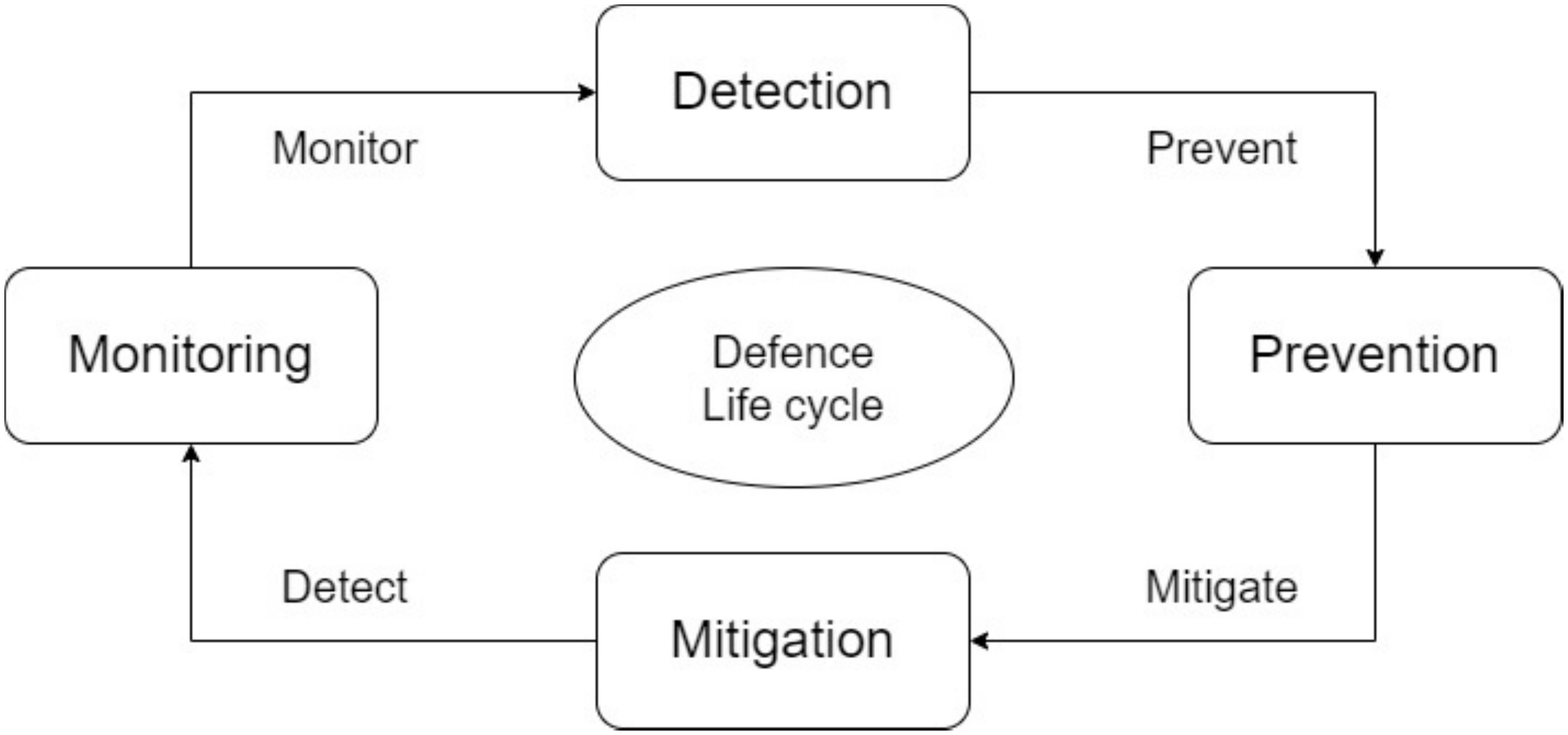}
\caption{Defence life cycle}
\label{FIG:Defence_life_cycle_converted}
\end{figure}

Other assaults include Land Attack, Mail Bomb, Ping of Death, Process Table, SSH Process Table, Syslogd, and so on. A DDoS on a protocol does not necessarily involve the characteristic of flooding by massive amounts of traffic. A well-known example would be the TCP SYN attack, which exploits the allocation of a connection context in the server\cite{article17}.To tackle these problems, defence mechanisms come into action. Fig. \ref{FIG:Defence_life_cycle_converted} illustrates, the DDoS defence life-cycle consists of four phases: prevention, monitoring, detection, and mitigation.

In the Prevention phase, to protect services and data from DDoS attacks, suitable security appliances are installed at various locations. In the monitoring phase, tools are used to collect important host or network information to track the system's execution. In the detection phase, tools are involved in the examination of the systems in use to identify the source of malicious traffic or DDoS attempts\cite{book18} \cite{article19}. Mitigation, the final phase of the defensive life-cycle, involves assessing the severity of the threat and determining the appropriate reaction at the appropriate moment. A response system determines appropriate countermeasures to successfully tackle a DDoS assault or slow down rogue clients during the mitigation phase\cite{article20}.

Existing DDoS defensive methods have had little success because they are unable to handle the significant challenge of providing efficient detection, effective reaction. Further, they are accompanied by unacceptable levels of false alarms, and real-time transmission of all packets all at the same time. Large financial losses have been incurred by top corporations throughout the world as a result of poor prevention and detection\cite{article21}. Prevention involves the implementation of a set of defences, practices, and configurations before any kind of DDoS attack, intending to reduce the impact of such an attack. Table \ref{TBL:Approaches_in_Prevention_of_DDoS} gives information about DDoS Prevention\cite{article22_29}.

\begin{table}
\centering
\caption{Approaches in Prevention of DDoS}
\label{TBL:Approaches_in_Prevention_of_DDoS}
\begin{tabular}{ p{4cm} p{7cm}   p{8cm}}
 \hline \hline
Approaches  &Description \\
 \hline

Over Provisioning   &This approach, which is based on preventing an attack on a site by preparing in advance for far more traffic than would be expected during normal operation, was the main prevention mechanism used, although now, with the increasing size of DDoS attacks, this is not sustainable. \\
Modifying Scheduling Algorithms &The goal of this strategy is to give legal traffic against malicious traffic. This approach incorporates the value of suspicion into its scheduling choice. The arrivals, request arrivals, and workload profiles sessions are used as input for the suspicion assignment method. \\
QoS and Resource Accounting &The goal of resource accounting is to keep track of resources across protected domains. It is useful in mitigating DDoS assaults since we need assured bandwidth, priority, and QoS during a DDoS attack. These resources are safeguarded by QoS methods like traffic shaping, scheduling, and congestion avoidance, which enforce traffic class priorities and provide fair service to genuine users. \\

\hline
\end{tabular}
\end{table}%

Furthermore, DDoS Detection is classified into two main categories, signature-based and behaviour-based. 
In Signature-based detection, the traffic is compared with the well-defined patterns of DDoS attacks that happened in history. This technique will be ineffective if the communication is encrypted\cite{article23}. And In Behaviour-based detection, is based on defining the standard to network traffic, so that any deviation can be considered malicious. 
The Behaviour is detected and standardized by

\ a. Comparing forward and reverse traffic.

\ The traffic is measured in both directions (uplink traffic and downlink traffic), it should be proportional, if not it is considered malicious. The drawback of this approach is that stealth attacks based on legitimate traffic cannot be detected\cite{article24}.

\ b. Aggregate traffic behavioural monitoring.

\ A network node's incoming or outgoing traffic is separated into multiple traffic aggregates and evaluated for trends. The intended communication should have the same characteristics, such as HTTP, encryption, or a UDP connection\cite{inproceedings25}.

\ c. Flow traffic behavioural monitoring.

\ The flow of traffic on the network is monitored, to enable fine-grained control of the behaviour of different flows\cite{inproceedings26}. The machine learning algorithms are used to learn the normal behaviour of network traffic and then detect abnormal behaviours.

On the other side, different defensive methods against attackers have been created \cite{article27} \cite{article28} \cite{article22_29}. Some well known existing DDoS Defence mechanisms are tabulated in Table \ref{TBL:Existing_Tools/Defensive mechanisms_in_defending DDoS}.

\begin{table}
\centering
\caption{Existing Tools/Defensive mechanisms in defending DDoS}
\label{TBL:Existing_Tools/Defensive mechanisms_in_defending DDoS}
\begin{tabular}{ p{5cm} p{7cm}   p{8cm}}
 \hline \hline
Defence mechanism &Description \\
 \hline
Protocol Security &Protocol security mechanisms address the problem of a bad protocol design. For example, many protocols contain operations that are cheap for the client but expensive for the server. \\
Filtering &Filtering mechanisms use the characterization provided by detection mechanisms to filter out the attack streams completely. Examples include dynamically deployed firewalls. \\
Resource multiplication &Resource multiplication mechanisms provide an abundance of resources to counter DDoS threats. In this, the system deploys a pool of servers with a load balancer and installs high bandwidth links between itself and upstream routers. \\
Network security &The victim network's DDoS defence systems shield the network against DDoS assaults and respond to identified attacks by reducing the impact on the victim. Most defensive systems are traditionally positioned at the victim, since it is the one that bears the brunt of the assault. \\
Disabling IP broadcast &The malicious part of this attack is that the attacker can use a low-bandwidth connection to destroy high bandwidth connections. The number of packets that are sent by the attacker is multiplied by a factor equal to the number of hosts behind the router that reply to the ICMP echo packets. \\

\hline
\end{tabular}
\end{table}%

\section{Existing Related Works and Advancement through the Current Paper}
\label{Sec:Related_Works}
To mitigate DDoS assaults, Hwang and Ku devised a distributed method. Distributed Change-point Detection (DCD) is a mitigation solution that largely minimizes the danger of such attacks. To discover any big or minor differences in network traffic, the researcher recommends utilizing the non-parametric CUSUM (Cumulative Sum) technique. For detection, the researchers also concentrated on the attack's original source. To prevent the assaults, Siaterlis \& Maglaris devised a technique based on single network features. The data was integrated with the traffic created by the experimenters using a data fusion method utilizing Multi-layer Perceptron (MLP), in which the inputs were initialized from different non-active measurements that were accessible on the network.

Joshi, Gupta, and Misra \cite{man32_42} employed a neural network design concept to discover zombie computers that were feeding DDoS assaults. The fundamental goal of their project was to figure out how the zombie computer and sample entropy were linked. With the aid of a feed-forward neural network, the entire process workflow is based on predictions. Another goal of their study is to use present infrastructure to identify and mitigate such assaults. To prevent DDoS assaults from disrupting network resources and services, Badishi, Yachin, and Keidar \cite{article33} employed cryptography and authentication. Shi, Stoica, and Anderson suggested a fairly similar strategy, although DDoS assaults are identified using a different technique termed perplexing mechanism.

Hari and Dohi \cite{article35} have published a paper on the importance of these protocol assaults. To address this, Badishi et al. \cite{article36} proposed a suggestion that restarts the protocol. By restarting the communication port number ordering on a regular basis, the sender, and receiver can use a fresh seed of pseudorandom functionality that creates varied port number arrangements. As a result, even if the attacker may launch a directed attack as a result of missing acknowledgement packets, the sender, and receiver can still interact simply by restarting the protocol. This restart is the result of the assumption that the difference in clock timings of two communication parties allocated to identify sender and recipient restart at the same time.

Badishi et al. \cite{article36} developed a detailed model and research on the subject of DoS to applications (ports) by an attacker who can listen in. Aside from the recommended port hopping protocols, this research will also look at the impact of adversaries' various techniques of launching blind assaults. The attacker's goal is to lower the delivery probability or the likelihood that client information will be collected by the server, as much as feasible. The researchers proposed a bottom constraint beyond which the attacker would be unable to diminish the delivery probability. The lower bound is dependent on a port's ability to receive messages as well as an attacker's ability to flood messages.Despite the application's protection mechanism, the results provide critical information about the settings and procedures discussed in this article.

Akilandeswari and Shalinie \cite{inproceedings37}, proposed a Probabilistic Neural Network-Based Attack Traffic taxonomy in order to detect various DDoS attacks. In contrast, the authors mainly focused on distinguishing between Flash Crowd Events from Denial of Service Attacks. Moreover, their work also involved the use of the Bayes decision rule for Bayes interference coupled with Radial Basis Function Neural Network (RBFNN) for precisely classifying the DDoS attack traffic and the legitimate traffic.To identify assaults, Liu, Gu, et al. developed a method called Learning Vector Quantization (LVQ) neural networks. The approach is a sort of quantization called supervision, and it may be used for things like pattern recognition, data compression, and multi-class classifications. In addition, the inputs to neural networks were provided as data sets in the form of numerical computations.

According to Vellalacheruvu and Kumar \cite{article39}, CP SYN flood is a typical DDoS assault, and contemporary operating systems provide a defence mechanism against this attack, which might impact web application performance and user connectivity. The performance of the TCP SYN attack defence that comes with Microsoft Windows Server 2003 is examined in this study. It has been discovered that the server's SYN attack security is only efficient at preventing attacks at low loads of SYN attack traffic and is insufficient to defend bigger magnitudes of SYN attack traffic. The study's units of findings can help network operators assess the protection mechanism existing in millions of Windows servers 2003 in defending the most common DDoS assault, known as the TCP SYN attack, and subsequently enhance their network security through other ways.

A group of authors \cite{snort40} \cite{man32_42} proposed a packet-marking and entropy system in which each packet is tagged on each router participating in communication to track the packet's origination. However, some writers presented a number of approaches that leveraged ANN or infrastructure to protect against DDoS assaults, while others detected the source of the attack. None of them, on the other hand, mentioned any unknown or zero-day threats classified as high or low risk. As a result, our primary goal is to identify unknown DDoS attempts.

\section{DDoSDet: The Proposed Method}
\label{Sec:DDoSDet}
\subsection{Preparation Of Dataset}
One of the broadest difficulties for ML/DL interruption recognition approaches is the accessibility of the datasets. The primary justification behind the absence of datasets in the interruption discovery area gets back to protection and unlawful issues. The organization traffic contains extremely delicate data, where the accessibility of such data can uncover clients and business insider facts, or even the individual correspondence. To cover the past hole, numerous specialists re-enact their own information to stay away from any touchy worries. Anyway, in these circumstances, the majority of the datasets produced are not extensive, and the column tests considered are not adequate to cover the application practices. The most well known public datasets, which have been utilized widely for interruption location are KDDCUP99 \cite{unknown43}, NSL-KDD \cite{article44}, Kyoto 2006+ \cite{article45}, ISCX2012 \cite{article46}, and CICIDS2017 \cite{article47}. More insights concerning different datasets utilized for interruption space can be found in \cite{article48}.

In this paper, we assess our proposed classifier utilizing the new delivered CICDDoS2019 dataset \cite{article49}. The dataset contains a lot of various DDoS assaults that can be helped out through application layer conventions utilizing TCP/UDP. The scientific categorization of assaults in the dataset is acted as far as abuse based and reflection-based assaults. The dataset was gathered in two separate days for preparing and testing assessment. The preparation set was caught on January twelfth, 2019, and contains 12 various types of DDoS assaults, each assault-type in an isolated PCAP document.

\begin{figure}
\centering
\includegraphics[width=0.5\textwidth]{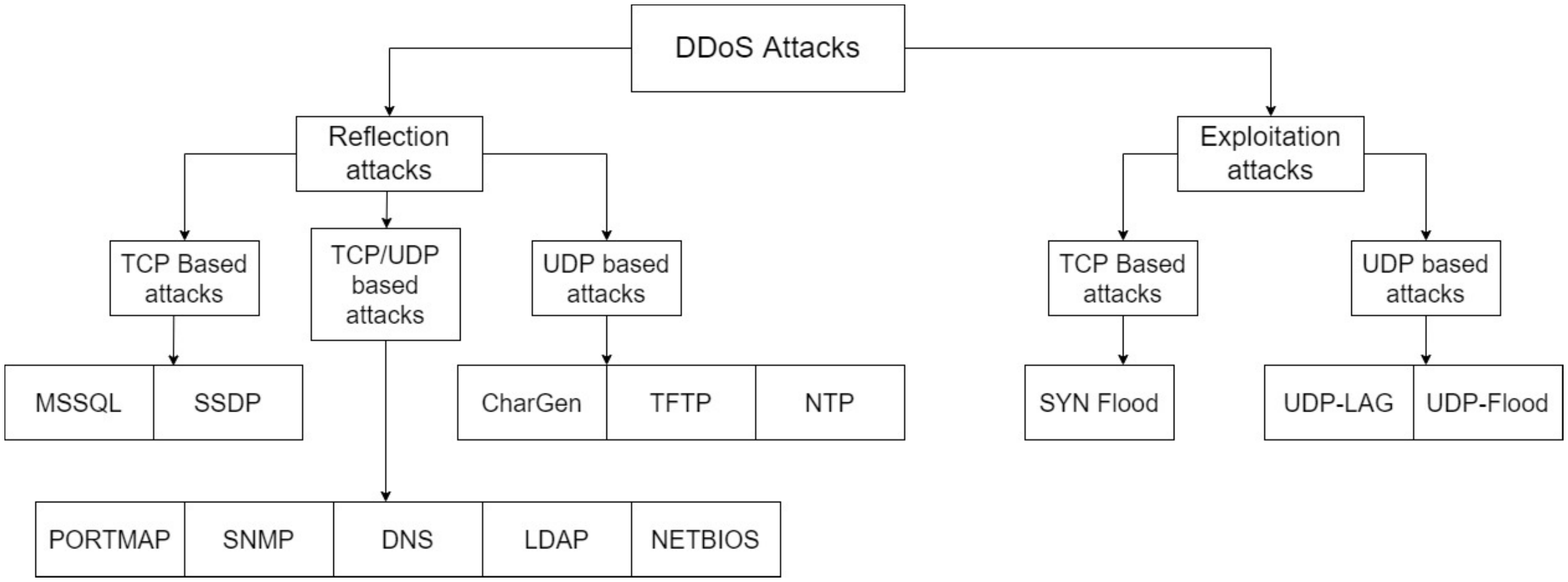}
\caption{The distribution of DoS and DDoS attacks inside the CICDDoS2019 dataset.}
\label{FIG:Flowchart_converted}
\end{figure}

The assault types in the preparation day incorporate UDP, SNMP, NetBIOS, LDAP, TFTP, NTP, SYN, WebDDoS, MSSQL, UDP-Lag, DNS, and SSDP DDoS based assaults. The testing information was made on March eleventh, 2019, and contains 7 DDoS assaults SYN, MSSQL, UDP-Lag, LDAP, UDP, PortScan, and NetBIOS. The dispersion of the various assaults in the dataset is displayed in Fig. \ref{FIG:Flowchart_converted}. The PortScan assault in the testing set is absent in the preparation information for the natural assessment of the location framework. The dataset contains in excess of 80 streams includes and was extricated utilizing CICFlowMeter apparatuses \cite{article50}. The CICDDoS2019 dataset is accessible on the site of the Canadian Institute for Cybersecurity in both PCAP record and stream design-based. We prepared the data to be suitable for the training model directly. CICDDoS2019 dataset is available in a flow-based format, where more than 80 features are extracted using CICFlowMeter.

\subsection{Network Architecture}
In the present era, Artificial Neural Networks (ANN) have gained a lot of popularity regarding their usage in real life. Artificial neural networks are used in Speech recognition. To ease the communication barrier, a simple solution could be, communication in a spoken language that is possible for the machine to understand with the help of ANN. This is mainly used in Virtual Assistants like Siri, Alexa and Google. Human Facial Recognition is one of the biometric approaches for identifying a certain face. Because of this, the classification of "non-face" pictures had become a common task. However, if a neural network is sufficiently trained, we can use it very efficiently. We have done the same for detecting DDoS attacks by using the CICDDoS2019 dataset. 

\begin{figure}
\centering
\includegraphics[width=0.5\textwidth]{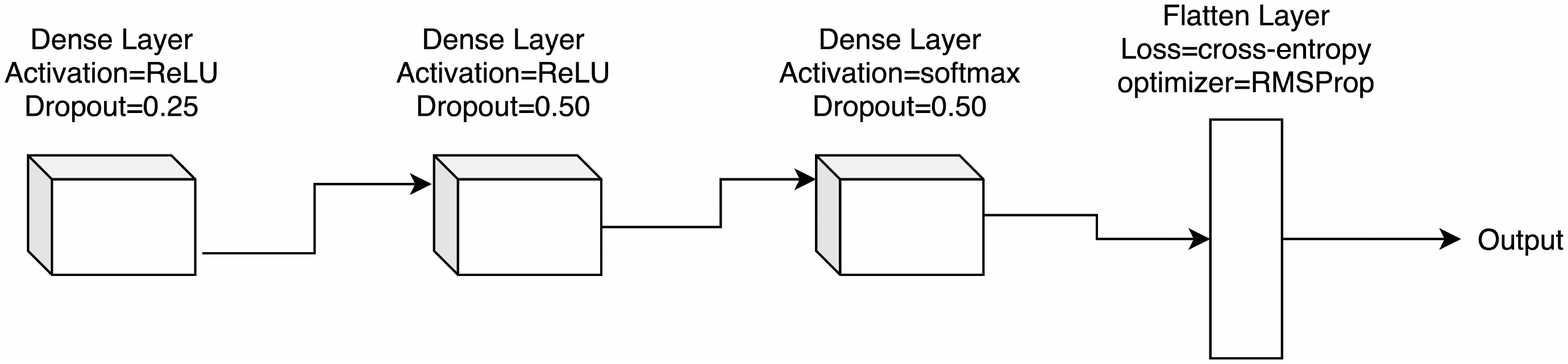}
\caption{Representation of Network Architecture.}
\label{FIG:network_arch-converted}
\end{figure}

Fig. \ref{FIG:network_arch-converted} presents the detailed network architecture. The base network of the model has input dimensions that equal the features of the dataset. This dense layer is processed by dropout of 0.25 with ReLU activation function, fed to a similar stack of 3 layers and compiled with categorical cross-entropy loss with "RMSProp" optimizer. The shape of every dense layer is 64 with the last flattened layer consisting of 2 outputs. The total trainable parameters of this sequential model are 9730.

\section{Results}
\label{Sec:results}
Our model achieved better results using a smaller value of dropout with a ReLU activation function for every layer. It can be observed that the model achieved overall accuracy up to 99.7\%, which was trained for 40 epochs. Furthermore, we changed the number of hidden layers, iteration, number of channels per hidden layer, and the activation function for each. The best performance is achieved when we used three hidden layers. When the number of hidden layers is increased, the model accuracy remains constant, but the training increases considerably. Therefore, we use three layers in our proposed framework. As a result, the three layers are more convincing to give reasonable results. Table \ref{TBL:Comparison_of_our_model_with_existing_algorithms} represents the performance evaluation of the proposed model with other classical ML algorithms. Our proposed approach performs best, as compared to the other bench marking algorithms.

\begin{figure}
\centering
\includegraphics[width=0.5\textwidth]{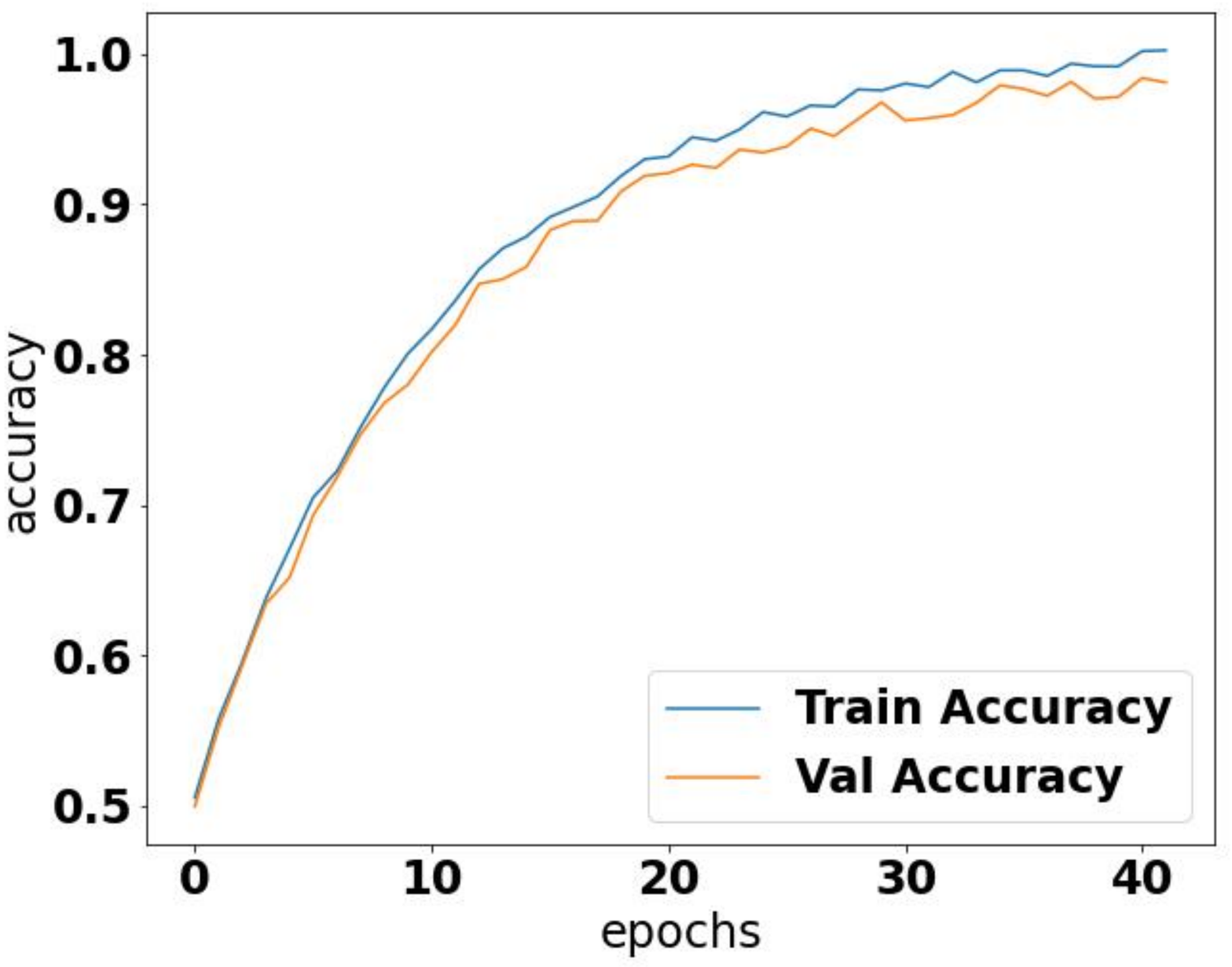}
\caption{Epochs vs Accuracy.}
\label{FIG:epochs_vs_acc-converted}
\end{figure}

\begin{figure}
\centering
\includegraphics[width=0.5\textwidth]{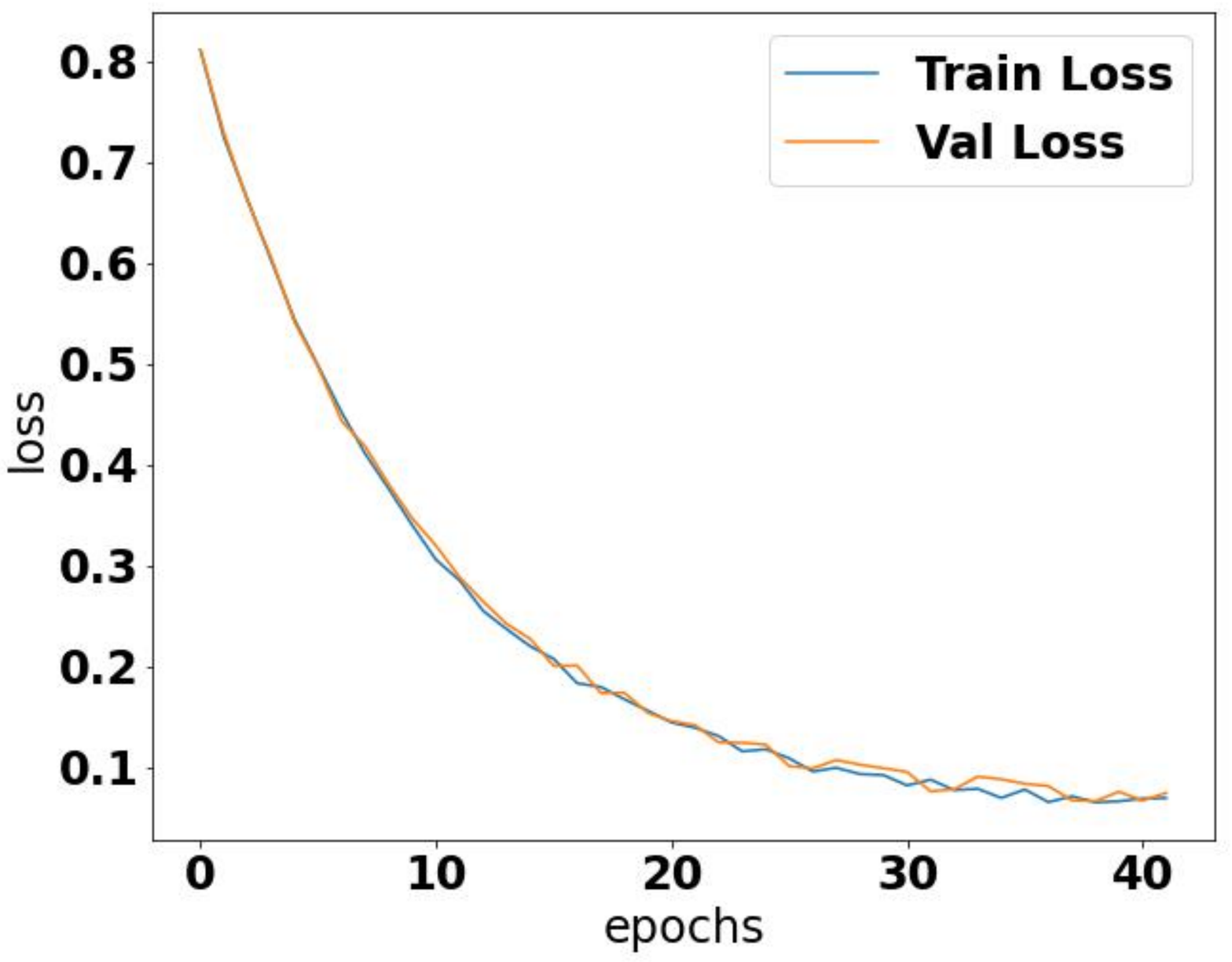}
\caption{Epochs vs Loss.}
\label{FIG:epochs_vs_loss-converted}
\end{figure}

\begin{table}
\centering
\caption{Comparison of our model with existing algorithms}
\label{TBL:Comparison_of_our_model_with_existing_algorithms}
\begin{tabular}{ p{2.10cm} p{2.50cm} p{2.50cm} p{2.50cm} p{2.30cm}}
 \hline \hline
Techniques  &Precision  &Recall  &F-score  &Accuracy(\%) \\
 \hline
 &Attack \& Benign &Attack \& Benign &Attack \& Benign & \\
 NB &1.00,  0.53 &0.17,  1.00 &0.29,  0.69 &57 \\
 DT &0.70,  0.98 &0.99,  0.54 &0.82,  0.70 &77 \\
 Booster &0.76,  0.99 &0.99,  0.67 &0.86,  0.80 &84 \\
 RF &1.00,  0.78 &0.74,  1.00 &0.85,  0.88 &86 \\
 SVM &0.99,  0.88 &0.88,  0.99 &0.93,  0.93 &93 \\
 LR &0.93,  0.99 &0.99,  0.91 &0.96,  0.95 &95 \\
 DDoSDet &0.97,  0.98 &0.99,  0.99 &0.98,  0.99 &99 \\
\hline
\end{tabular}
\end{table}%

The performance of the classification methods not only depends on the used technique, but also on the manner in which training and testing data is partitioned. We used samples from each attack type to obtain a balanced dataset with respect to different types of attacks. In our case, the total number of samples for training and validation sets are 271479 and 53880, respectively. To get a realistic detection rate, we used attack records in the testing set that are not represented in the training phase. The total number of records in the testing set is 28000 samples. Fig. \ref{FIG:epochs_vs_acc-converted} and \ref{FIG:epochs_vs_loss-converted} represent the epochs vs accuracy and epochs vs loss respectively.

\section{Conclusion}
\label{Sec:conclusion}

Network virtualization prompts new dangers and new exploitable assaults than the ones previously existed in the conventional organization. The DDoS assault class is viewed as one of the most forceful assault types lately, causing a basic effect all in all network frameworks. In this paper, we proposed a model DDoSDet that depends on neural networks for the detection of DDoS assaults against SDN organizations. We utilized the new delivered CICDDoS2019 dataset for preparing and assessment of our proposed model. The dataset contains far-reaching and latest DDoS sorts of assaults. The assessment of our model showed that DDoSDet gives the most elevated assessment measurements as far as recall, precision, F-score, and accuracy contrasted with the current notable old-style machine learning methods.

\section*{Acknowledgments} 

This research is carried in the Artificial Intelligence and Robotics (AIR) Research Centre, VIT-AP University. We also thank the management for motivating and supporting AIR Research Centre, VIT-AP University in building this project.

\bibliographystyle{unsrtnat}
\bibliography{Arxiv_30.10.2021-EfficientWord-Net}

\begin{thebibliography}{40}
\providecommand{\natexlab}[1]{#1}
\providecommand{\url}[1]{\texttt{#1}}
\expandafter\ifx\csname urlstyle\endcsname\relax
  \providecommand{\doi}[1]{doi: #1}\else
  \providecommand{\doi}{doi: \begingroup \urlstyle{rm}\Url}\fi

\bibitem[Ahamad and Aljumah(2015)]{article1}
Tariq Ahamad and Abdullah Aljumah.
\newblock Detection and defense mechanism against ddos in manet.
\newblock \emph{Indian Journal of Science and Technology}, 8, 12 2015.
\newblock \doi{10.17485/ijst/2015/v8i33/80152}.

\bibitem[wha()]{whatiscybersecurity?}
URL
  \url{https://www.cisco.com/c/en_in/products/security/what-is-cybersecurity.html}.

\bibitem[ddo()]{ddos1}
URL
  \url{https://www.paloaltonetworks.com/cyberpedia/what-is-a-denial-of-service-attack-dos}.

\bibitem[Aldaej and Ahamad(2016)]{article2}
Abdulaziz Aldaej and Tariq Ahamad.
\newblock Aaodv (aggrandized ad hoc on demand vector): A detection and
  prevention technique for manets.
\newblock \emph{International Journal of Advanced Computer Science and
  Applications}, 7, 10 2016.
\newblock \doi{10.14569/IJACSA.2016.071018}.

\bibitem[ids()]{ids1}
URL
  \url{https://www.techtarget.com/searchsecurity/definition/intrusion-detection-system}.

\bibitem[Perez~del Pino et~al.(2010)Perez~del Pino, Baez, Lopez, and
  Suarez~Araujo]{inproceedings1}
Miguel~Angel Perez~del Pino, P.~Baez, P.~Lopez, and Carmen Suarez~Araujo.
\newblock Towards self-organizing maps based computational intelligent system
  for denial of service attacks detection.
\newblock pages 151 -- 157, 06 2010.
\newblock \doi{10.1109/INES.2010.5483858}.

\bibitem[Lee et~al.(2003)Lee, Choi, Jung, and Noh]{article8}
Cheo-Iho Lee, Kyung~Hee Choi, Gi~Hyun Jung, and Sang~Guk Noh.
\newblock An analysis of network traffic on ddos attacks against web servers.
\newblock \emph{The KIPS Transactions:PartC}, 10C, 06 2003.
\newblock \doi{10.3745/KIPSTC.2003.10C.3.253}.

\bibitem[Junior and Kumar(2014)]{article9}
Rodolfo Junior and Sanjeev Kumar.
\newblock Apple's lion vs microsoft's windows 7: Comparing built-in protection
  against icmp flood attacks.
\newblock \emph{Journal of Information Security}, 05:\penalty0 123--135, 01
  2014.
\newblock \doi{10.4236/jis.2014.53012}.

\bibitem[Canada()]{canada11}
Bell Canada.
\newblock Bell blog.
\newblock URL
  \url{https://blog.bell.ca/costs-and-consequences-of-a-distributed-denial-of-service-ddos-attack/}.

\bibitem[Fitri(2021)]{tech13}
Afiq Fitri.
\newblock Navigating the horizon of business technology, Mar 2021.
\newblock URL \url{http://www.cbronline.com/news/security/internet-slows-d}.

\bibitem[Kitten and Ross()]{kitten14}
Tracy Kitten and Ron Ross.
\newblock 2 more banks are ddos victims.
\newblock URL
  \url{https://www.bankinfosecurity.com/2-more-banks-are-ddos-victims-a-5298}.

\bibitem[Tripathi et~al.(2013)Tripathi, Gupta, Almomani, Mishra, and
  Veluru]{article15}
Shweta Tripathi, B~B Gupta, Dr.Ammar Almomani, Anupama Mishra, and Suresh
  Veluru.
\newblock Hadoop based defense solution to handle distributed denial of service
  (ddos) attacks.
\newblock \emph{Journal of Information Security}, 04:\penalty0 150--164, 01
  2013.
\newblock \doi{10.4236/jis.2013.43018}.

\bibitem[the(2021)]{the16}
Microsoft said it mitigated a 2.4 tbps ddos attack, Oct 2021.

\bibitem[Moore et~al.(2006)Moore, Shannon, Brown, Voelker, and
  Savage]{article17}
David Moore, Colleen Shannon, Douglas Brown, Geoffrey Voelker, and Stefan
  Savage.
\newblock Inferring internet denial-of-service activity.
\newblock \emph{ACM Trans. Comput. Syst.}, 24:\penalty0 115--139, 05 2006.
\newblock \doi{10.1145/1132026.1132027}.

\bibitem[Mirkovic et~al.(2004)Mirkovic, Dietrich, Dittrich, and Reiher]{book18}
Jelena Mirkovic, Sven Dietrich, David Dittrich, and Peter Reiher.
\newblock \emph{Internet Denial of Service: Attack and Defense Mechanisms
  (Radia Perlman Computer Networking and Security Book Series)}.
\newblock 12 2004.
\newblock ISBN 0131475738.

\bibitem[Yu et~al.(2014)Yu, Tian, Guo, and Wu]{article19}
Shui Yu, Yonghong Tian, Song Guo, and Dapeng Wu.
\newblock Can we beat ddos attacks in clouds?
\newblock \emph{IEEE Transactions on Parallel and Distributed Systems}, 25, 09
  2014.
\newblock \doi{10.1109/TPDS.2013.181}.

\bibitem[Chen(2010)]{article20}
P.M. Chen.
\newblock Acm transactions on computer systems: Editorial.
\newblock 28, 03 2010.
\newblock \doi{10.1145/1731060.1731061}.

\bibitem[Alomari et~al.(2012)Alomari, Manickam, Gupta, Karuppayah, and
  Alfaris]{article21}
Esraa Alomari, Selvakumar Manickam, B~B Gupta, Shankar Karuppayah, and Rafeef
  Alfaris.
\newblock Botnet-based distributed denial of service (ddos) attacks on web
  servers: Classification and art.
\newblock \emph{International Journal of Computer Applications}, 49, 08 2012.
\newblock \doi{10.5120/7640-0724}.

\bibitem[Shameli-Sendi et~al.(2015)Shameli-Sendi, Pourzandi, Fekih-Ahmed, and
  Cheriet]{article22_29}
Alireza Shameli-Sendi, Makan Pourzandi, Mohamed Fekih-Ahmed, and Mohamed
  Cheriet.
\newblock Taxonomy of distributed denial of service mitigation approaches for
  cloud computing.
\newblock \emph{Journal of Network and Computer Applications}, 58, 10 2015.
\newblock \doi{10.1016/j.jnca.2015.09.005}.

\bibitem[Peng et~al.(2007)Peng, Leckie, and Ramamohanarao]{article23}
Tao Peng, Christopher Leckie, and Kotagiri Ramamohanarao.
\newblock Survey of network-based defense mechanisms countering the dos and
  ddos problems.
\newblock \emph{ACM Comput. Surv.}, 39, 04 2007.
\newblock \doi{10.1145/1216370.1216373}.

\bibitem[Gil and Poletto(2001)]{article24}
Thomer Gil and Massimiliano Poletto.
\newblock Multops: A data-structure for bandwidth attack detection.
\newblock 06 2001.

\bibitem[Kuzmanovic and Knightly(2003)]{inproceedings25}
Aleksandar Kuzmanovic and Edward Knightly.
\newblock Low-rate tcp-targeted denial of service attacks: the shrew vs. the
  mice and elephants.
\newblock pages 75--86, 10 2003.
\newblock \doi{10.1145/863955.863966}.

\bibitem[Bernaille and Teixeira(2007)]{inproceedings26}
Laurent Bernaille and Renata Teixeira.
\newblock Early recognition of encrypted applications.
\newblock pages 165--175, 04 2007.
\newblock ISBN 978-3-540-71616-7.
\newblock \doi{10.1007/978-3-540-71617-4_17}.

\bibitem[Mirkovic and Reiher(2004)]{article27}
Jelena Mirkovic and Peter Reiher.
\newblock A taxonomy of ddos attack and ddos defense mechanisms.
\newblock \emph{ACM SIGCOMM Computer Communication Review}, 34, 05 2004.
\newblock \doi{10.1145/997150.997156}.

\bibitem[Aljumah(2017)]{article28}
Abdullah Aljumah.
\newblock Detection of distributed denial of service attacks using artificial
  neural networks.
\newblock \emph{International Journal of Advanced Computer Science and
  Applications}, 8, 01 2017.
\newblock \doi{10.14569/IJACSA.2017.080841}.

\bibitem[man()]{man32_42}
Iptables.
\newblock URL \url{http://ipset.netfilter.org/iptables.man.html}.

\bibitem[Leu and Pai(2009)]{article33}
Fang-Yie Leu and Chia-Chi Pai.
\newblock Detecting dos and ddos attacks using chi-square.
\newblock In \emph{2009 Fifth International Conference on Information Assurance
  and Security}, volume~2, pages 255--258, 2009.
\newblock \doi{10.1109/IAS.2009.292}.

\bibitem[Hari and Dohi(2010)]{article35}
Kousaburou Hari and Tadashi Dohi.
\newblock Sensitivity analysis of random port hopping.
\newblock In \emph{2010 7th International Conference on Ubiquitous Intelligence
  Computing and 7th International Conference on Autonomic Trusted Computing},
  pages 316--321, 2010.
\newblock \doi{10.1109/UIC-ATC.2010.69}.

\bibitem[Badishi et~al.(2007)Badishi, Herzberg, and Keidar]{article36}
Gal Badishi, Amir Herzberg, and Idit Keidar.
\newblock Keeping denial-of-service attackers in the dark.
\newblock \emph{Dependable and Secure Computing, IEEE Transactions on},
  4:\penalty0 191--204, 08 2007.
\newblock \doi{10.1109/TDSC.2007.70209}.

\bibitem[Senthil~Kumar and Shalinie(2012)]{inproceedings37}
V.~Akilandeswari Senthil~Kumar and S.~Shalinie.
\newblock Probabilistic neural network based attack traffic classification.
\newblock 12 2012.
\newblock \doi{10.1109/ICoAC.2012.6416848}.

\bibitem[Vellalacheruvu and Kumar(2011)]{article39}
Hari Vellalacheruvu and Sanjeev Kumar.
\newblock Effectiveness of built-in security protection of microsoft's windows
  server 2003 against tcp syn based ddos attacks.
\newblock \emph{J. Information Security}, 2:\penalty0 131--138, 01 2011.
\newblock \doi{10.4236/jis.2011.23013}.

\bibitem[sno()]{snort40}
Ai.
\newblock URL \url{http://snort-ai.sourceforge.net/index.php}.

\bibitem[Parekh et~al.(2018)Parekh, Savla, Mishra, and Shirole]{unknown43}
Meet Parekh, Vaibhav Savla, Rudra Mishra, and Mahesh Shirole.
\newblock Benchmarking datasets for anomaly-based network intrusion detection:
  Kdd cup 99 alternatives, 10 2018.

\bibitem[Tavallaee et~al.(2009)Tavallaee, Bagheri, Lu, and Ghorbani]{article44}
Mahbod Tavallaee, Ebrahim Bagheri, Wei Lu, and Ali~A. Ghorbani.
\newblock A detailed analysis of the kdd cup 99 data set.
\newblock In \emph{2009 IEEE Symposium on Computational Intelligence for
  Security and Defense Applications}, pages 1--6, 2009.
\newblock \doi{10.1109/CISDA.2009.5356528}.

\bibitem[art()]{article45}
Description of kyoto university benchmark data - takakura.com.
\newblock URL
  \url{https://www.takakura.com/Kyoto_data/BenchmarkData-Description-v5.pdf}.

\bibitem[Shiravi et~al.(2012)Shiravi, Shiravi, Tavallaee, and
  Ghorbani]{article46}
Ali Shiravi, Hadi Shiravi, Mahbod Tavallaee, and Ali Ghorbani.
\newblock Toward developing a systematic approach to generate benchmark
  datasets for intrusion detection.
\newblock \emph{Computers \& Security}, 31:\penalty0 357–374, 05 2012.
\newblock \doi{10.1016/j.cose.2011.12.012}.

\bibitem[Sharafaldin et~al.(2018)Sharafaldin, Habibi~Lashkari, and
  Ghorbani]{article47}
Iman Sharafaldin, Arash Habibi~Lashkari, and Ali Ghorbani.
\newblock Toward generating a new intrusion detection dataset and intrusion
  traffic characterization.
\newblock pages 108--116, 01 2018.
\newblock \doi{10.5220/0006639801080116}.

\bibitem[Ring et~al.(2019)Ring, Wunderlich, Scheuring, Landes, and
  Hotho]{article48}
Markus Ring, Sarah Wunderlich, Deniz Scheuring, Dieter Landes, and Andreas
  Hotho.
\newblock A survey of network-based intrusion detection data sets.
\newblock \emph{Computers \& Security}, 86, 06 2019.
\newblock \doi{10.1016/j.cose.2019.06.005}.

\bibitem[Sharafaldin et~al.(2019)Sharafaldin, Habibi~Lashkari, Hakak, and
  Ghorbani]{article49}
Iman Sharafaldin, Arash Habibi~Lashkari, Saqib Hakak, and Ali Ghorbani.
\newblock Developing realistic distributed denial of service (ddos) attack
  dataset and taxonomy.
\newblock pages 1--8, 10 2019.
\newblock \doi{10.1109/CCST.2019.8888419}.

\bibitem[Habibi~Lashkari et~al.(2016)Habibi~Lashkari, Draper~Gil, Mamun, and
  Ghorbani]{article50}
Arash Habibi~Lashkari, Gerard Draper~Gil, Mohammad Mamun, and Ali Ghorbani.
\newblock Characterization of encrypted and vpn traffic using time-related
  features.
\newblock 02 2016.
\newblock \doi{10.5220/0005740704070414}.

\end{thebibliography}

\end{document}